\documentclass[prl,twocolumn,showpacs,preprintnumbers,amsmath,amssymb]{revtex4}

\usepackage{graphicx}
\usepackage{dcolumn}
\usepackage{bm}

\begin{document}


\title{Ferromagnetism in the Mott insulator Ba$_2$NaOsO$_6$}

\author{A. S.  Erickson,$^1$ S. Misra,$^2$ G. J. Miller,$^2$ R. R. Gupta,$^3$ Z. Schlesinger,$^3$ W. A. Harrison,$^1$ J. M. Kim,$^1$ I. R. Fisher$^1$} 
\affiliation{$^1$Department of Applied Physics and Geballe Laboratory for Advanced Materials, Stanford University, CA 94305.\\
$^2$Department of Chemistry and Ames Laboratory, Iowa State University, Ames, IA 50011-2300\\$^3$Department of Physics, University of California, Santa Cruz, CA 95064}

\date{\today}

\begin{abstract}
Results are presented of single crystal structural, thermodynamic, and 
reflectivity measurements
of the double-perovskite Ba$_2$NaOsO$_6$. These characterize the material as a 5$d^1$ ferromagnetic Mott insulator with an ordered moment of 
$\sim$ 0.2 $\mu_B$ per formula unit and $T_C=$ 6.8(3) K.  The magnetic entropy associated with this phase transition is close to $R$ln2, 
indicating that the quartet groundstate anticipated from consideration of the crystal structure is split, consistent with a scenario in 
which the ferromagnetism is associated with orbital ordering.  

\end{abstract}

\maketitle

The interplay between spin, orbital and charge degrees of freedom in 3$d$ transition metal oxides has proven to be a rich area of research in recent years. 
Despite the wide array of interesting physics found in these materials, much less is known about whether similar 
behavior can be found in related 4$d$ and 5$d$ systems, for which both the extent of the $d$-orbitals 
and larger spin-orbit coupling cause a different balance between the relevant energy scales. 
In this respect, oxides of osmium are of particular interest because the element can take formal 
valences from 4+ to 7+, corresponding to electron configurations 5$d^4$ to 5$d^1$. 
In this instance, we examine the simplest case of a 5$d^1$ osmate for which the magnetic properties 
indicate that orbital ordering may indeed play a significant role.  

Simple oxides of osmium are typically Pauli paramagnets due to the large extent of the $5d$ orbitals.  
Examples include the binary oxide OsO$_2$ \cite{oso2, oso2b} and the simple perovskites $A$OsO$_3$ ($A$= Sr, Ba) \cite{chamberland}. 
However, more complex oxides, 
including the double and triple perovskites La$_2$NaOsO$_6$ \cite{La2NaOsO6}, 
Ba$_2A$OsO$_6$ ($A$ = Li, Na) \cite{zl02,sleight} and Ba$_3A$Os$_2$O$_9$ (A = Li, Na) \cite{zl03},  appear to exhibit local moment behavior. 
Presumably the large separation of Os ions in these more complex structures leads to a Mott insulating state, 
and indeed these and related materials are most often found to be antiferromagnetic. 
Of the above materials and their near relations containing no other magnetic ions, 
Ba$_2$NaOsO$_6$ distinguishes itself as the only osmate with a substantial \textit{ferromagnetic} moment ($\sim$0.2 $\mu_B$) 
in the ordered state \cite{zl02}.  

Weak ferromagnetism has been previously observed in other $5d$ transition metal oxides containing iridium. BaIrO$_3$ exhibits a saturated moment of 0.03 $\mu_B$, which has been attributed to small exchange splitting associated with charge density wave formation \cite{Brooks2005}. Sr$_2$IrO$_4$ and Sr$_3$Ir$_2$O$_7$ exhibit similarly small saturated moments, 
attributed variously to either spin canting in an antiferromagnet due to the low crystal symmetry \cite{IrCrawford} or to a borderline metallic Stoner scenario \cite{Cao1998,Cao2002}. The ferromagnetic moment in Ba$_2$NaOsO$_6$ is substantially larger than in these materials. 
Furthermore, at room temperature the material has an undistorted double-perovskite structure,  space group Fm$\bar{3}$m (inset to Fig. 1) \cite{zl02}, in which OsO$_6$ octahedra are neither distorted nor rotated with respect to each other or the underlying lattice \cite{tolerance_factor}. 
Such a high crystal symmetry, if preserved to low temperatures, precludes the more usual mechanisms for obtaining a small ferromagnetic moment 
in an insulating antiferromagnet \cite{Moriya}, 
suggesting that a different mechanism is causing the ferromagnetism. 

\begin{figure}
\includegraphics[scale=0.55]{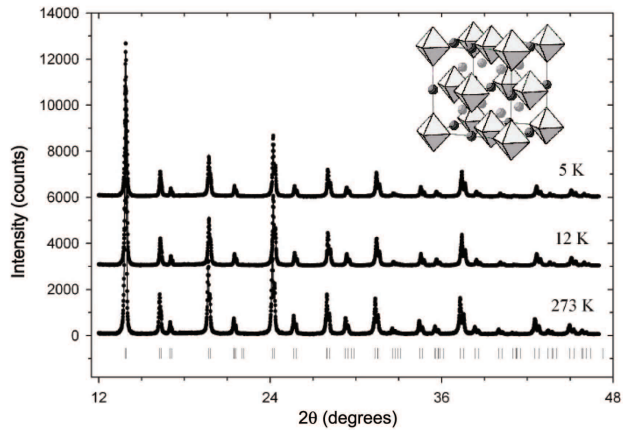}
\caption{\label{xrd}Observed (dots) and calculated (line) X-ray powder diffraction patterns of Ba$_2$NaOsO$_6$ collected at 273, 12 and 5 K. Vertical bars under the observed and calculated diffraction patterns indicate calculated positions of Bragg peaks.  Inset depicts the refined crystal structure, showing OsO$_6$ octahedra, Ba (light gray) and Na (dark gray) atoms.}
\end{figure}

Black, shiny, single crystals of Ba$_2$NaOsO$_6$ up to 2 mm in diameter and with a truncated octahedral morphlogy 
were grown from a molten hydroxide flux following a method similar to that presented in reference ~\onlinecite{zl02}. 
Single crystal x-ray diffraction data were collected at room temperature 
using both a STOE Image Plate Diffractometer (IPDS II) 
and a Bruker Smart Apex CCD diffractometer. 
Data were taken for a crystal with dimensions $0.23 \times 0.18 \times 0.01$ mm$^3$, 
and for a smaller piece broken from this larger crystal with dimensions $0.058 \times 0.039 \times 0.022$ mm$^3$, more closely approximating a sphere. 
In each case, a large number of reflections were collected (3074 and 2166 respectively), 
and the structure was refined using the SHELXTL package of crystallographic programs \cite{shellXRD}. 
Refinement to the published fully occupied, stoichiometric Fm$\bar{3}$m structure \cite{zl02} 
consistently gave the lowest R values (0.0147) against a number of variable parameters, 
including partial occupancy, mixed site occupancy and lower space group symmetry. 
These measurements were repeated at temperatures of 243, 223, 198, and 183 K, with no difference in the refined structures. 
Additional powder diffraction measurements were taken at temperatures of 273, 30, 12 and 5 K (Fig. \ref{xrd}), 
using a Rigaku TTRAX powder diffractometer, equipped with a helium-flow cryostat. 
The structure obtained by Rietveld refinement using the Rietica LHPM software \cite{PWDXRD} 
agreed with the single crystal structural refinement at all temperatures.  
Goodness of fit parameters were constant through the magnetic transition, indicating no discernable change in the crystal structure.

\begin{figure}
\includegraphics[scale=0.47]{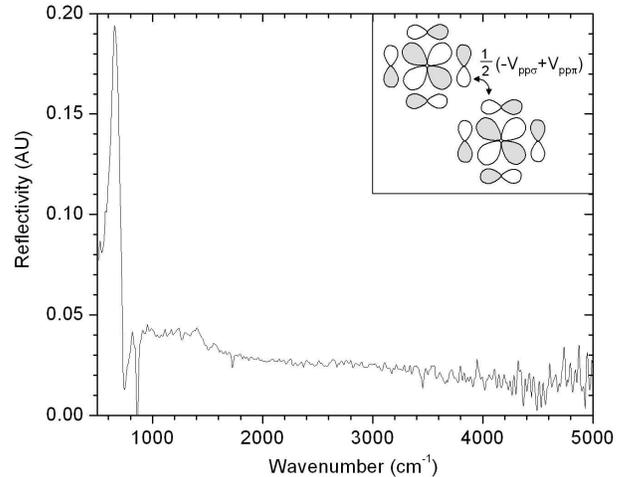}
\caption{\label{ir}IR reflectivity data as a function of wavenumber for Ba$_2$NaOsO$_6$.  Inset shows coupling between two adjacent OsO$_6$ orbitals of $d_{xy}$ symmetry for $k$ $=$ 0.}
\end{figure}

With one electron per osmium site, one might naively expect that Ba$_2$NaOsO$_6$ would be a metal. DC resistivity measurements consistently showed insulating behavior, but concern over the quality of the electrical contact to the samples led us to verify this observation by infrared reflectivity.
Measurements were carried out using a scanning Fourier transform
interferometer with a bolometer detector at 4.2 K, 
for arbitrary crystal orientations with the sample held at room temperature (Fig. \ref{ir}). 
An evaporated Ag film, 
adjacent to and co-planar with the sample, provided a reference used to
obtain absolute reflectivity versus frequency. 
The data show low overall reflectivity, with no indication of a metallic plasma edge down to the lowest measured frequency of 400 cm$^{-1}$. 
The strong variations in the reflectivity between 400 and 1000 cm$^{-1}$ are a signature of 
unscreened optical phonons. These data suggest that Ba$_2$NaOsO$_6$ is a Mott insulator, which is supported by the following tight-binding analysis. 

The crystal structure of Ba$_2$NaOsO$_6$ can be pictured as an FCC lattice of isolated OsO$_6$ octahedra separated by Ba and Na ions (inset to Fig \ref{xrd}).  
The $s$ orbitals of Ba and Na are so high in energy that they can be neglected, 
and the electronic structure is primarily determined by the OsO$_6$ octahedra, which form the usual set of molecular orbitals. 
The bonding and non-bonding states are filled, leaving one electron in the triply-degenerate $t_{2g}$ antibonding orbitals. 
Using the known energies for the Os and O orbitals we find that this molecular orbital 
is at -13.43 eV \cite{walt}, 
relative to E$_p$(O) = -16.77 eV for oxygen 2$p$ states and 
E$_d$(Os) = -16.32 eV \cite{walt1}.
Since the Os $d$ and O $p$ orbitals are close in energy, these molecular orbitals have almost equal 5$d$ and 2$p$ character. 

Adjacent OsO$_6$ octahedra in Ba$_2$NaOsO$_6$ are coupled by the matrix elements 
$\frac{1}{2}$(-V$_{pp\sigma}$ + V$_{pp\pi}$) (inset to Fig. \ref{ir}). 
Using values obtained for similar cluster separations in other materials \cite{walt}, 
modified appropriately for this particular lattice, and neglecting spin-orbit coupling, 
we obtain $t = 1/4 \times 1/2(-V_{pp\sigma} + V_{pp\pi}) \sim$ 0.05 eV 
for the hopping matrix elements coupling adjacent octahedra. 
In contrast, the Coulomb energy associated with moving one electron from an OsO$_6$ octahedron to its neighbor is 
found to be $U \sim$ 3.3 eV \cite{walt2}. 
More detailed treatments could presumably refine these values, but since we find $U \gg t$ it is clear that the material is a Mott insulator and that a local moment description of the magnetism is appropriate.

\begin{figure}
\includegraphics[scale=0.9]{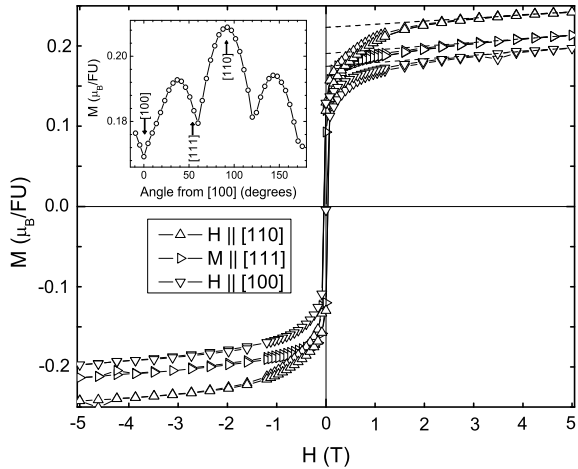} 
\caption{\label{mh}Magnetization of Ba$_2$NaOsO$_6$ along high symmetry directions as a function of applied field at 1.8 K for a full hysteresis loop. 
'FU' refers to one formula unit. Inset shows Magnetization as a function of angle in the (01$\bar{1}$) plane at a temperature of 1.8 K and a field of 2 T.  A line is drawn between data points to guide the eye.}
\end{figure}

Magnetization measurements as a function of applied field at 1.8 K (Fig. \ref{mh}) 
show ferromagnetic behavior, 
as previously reported for polycrystalline samples \cite{zl02}. 
These data were obtained for applied fields oriented along high symmetry directions using a 
Quantum Design Superconducting Quantum Interference Device magnetometer for single crystals weighing between 2 and 7 mg.   
The magnetization rises rapidly in low fields and levels off for fields above 1 T, beyond which there is no discernable hysteresis. 
However, the absolute value of the magnetization at this field is relatively small (approximately 0.2 $\mu_B$) and 
does not appear to saturate in fields of up to 5 T. Extrapolating a linear fit to the magnetization between 3 and 5 T, 
we obtain a zero-field magnetization of 0.175(2), 0.191(1), and 0.223(1) $\mu_B$ for fields oriented along [100], [111], and [110], respectively. 
This anisotropy is confirmed by angle-dependent measurements at a constant field using a rotating sample holder and
rotating in the (01$\bar{1}$) plane, which contains all three high-symmetry directions (inset to Fig. \ref{mh}).

\begin{figure}
\centering
\includegraphics[scale=0.95]{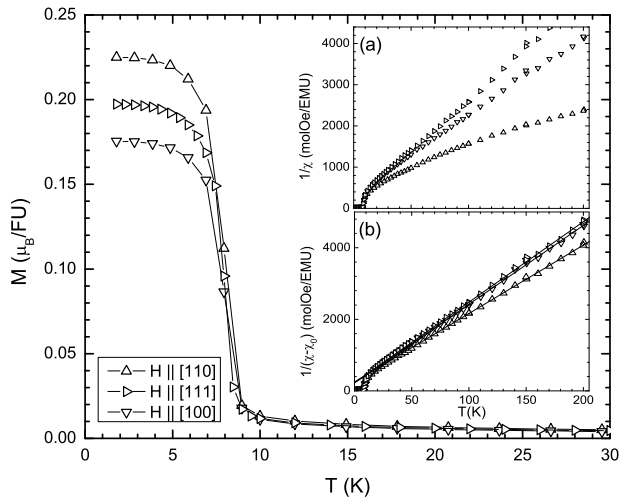}
\caption{\label{mt}Low temperature magnetization of Ba$_2$NaOsO$_6$ as a function of temperature in a field of 2 T. Insets show (a) inverse susceptibility and (b) the inverse of the susceptibility  with a constant offset, $\chi_0$, subtracted. Lines show fits to Curie-Weiss behavior. 'FU' refers to one formula unit.}
\end{figure}

Temperature-dependent magnetization measurements in fields above 1 T, for which there is no hysteresis, 
show an upturn below approximately 8 K (Fig. \ref{mt}), 
consistent with ferromagnetic behavior. 
The inverse susceptibility at high temperature shows a moderate amount of curvature (panel (a) of the inset to Fig. \ref{mt}). 
Data between 75 and 200 K can nevertheless be fit by a Curie-Weiss law if a constant offset, nominally attributed to Van Vleck paramagnetism, is included: 
$\chi$ = C/(T-$\theta$) + $\chi_0$ 
(panel (b) of the inset to Fig. \ref{mt}). 
This fit results in relatively small effective moments $\mu_{eff}$ of 0.602(4), 0.596(1), and 0.647(3) $\mu_B$ 
for fields oriented in the [100], [111], and [110] directions respectively, indicative of substantial spin-orbit coupling, and 
Weiss temperatures of $-10(2)$, $-10(1)$, and $-13(1)$ K. 
Values of $\chi_0$ are found to be 3(1)$\times10^{-5}$, 1.7(1)$\times10^{-4}$, and 1.76(9)$\times10^{-4}$ emu/molOe 
for fields oriented in these same directions. 

\begin{figure}
\centering
\includegraphics[scale = 0.8]{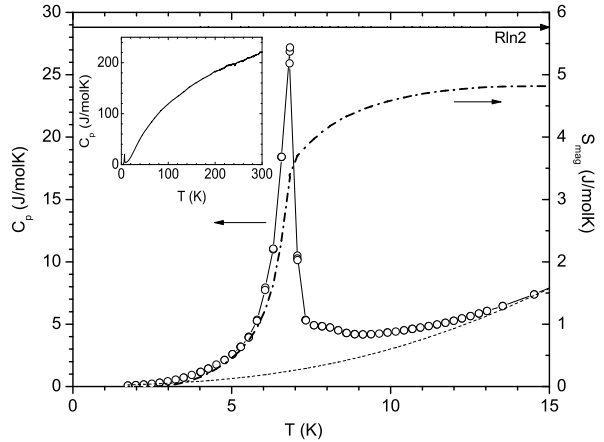}
\caption{\label{cp} Heat capacity (left axis) and magnetic contribution to the integrated entropy (right axis) for the magnetic transition in Ba$_2$NaOsO$_6$. The dashed line (left axis) shows an extrapolation of a fit to the phonon background, the solid horizontal line (right axis) indicates the theoretical entropy of $R$ln2 for a doublet groundstate. Inset shows the heat capacity to 300 K. Upper axis shows the Dulong-Petite value of 249 J/molK.}
\end{figure}

The magnetic phase transition is most clearly seen in the heat capacity (Fig. \ref{cp}). 
Data were taken using the relaxation method for 3-4 single crystals weighing a total of 2-3 mg, 
oriented at arbitrary angles to each other, for temperatures from 0.3 to 300 K. 
A sharp anomaly is seen with a peak at 6.8(3) K, which defines the critical temperature $T_C$. 
The magnetic contribution to the heat capacity was estimated by subtracting a polynomial extrapolation 
of the higher temperature phonon heat capacity, fit between 14.5 and 18 K (C$_{ph}$ = $0.065(7)T+0.00262(6)T^3-2.6(1)\times10^{-6}T^5$, 
shown in Fig. \ref{cp}). 
Since the magnetic contribution is significantly larger close to $T_C$, this crude subtraction results in only small systematic errors. The total magnetic contribution to the entropy $S_{mag}$ 
through the transition is found to be 4.6 J/molK (right axis, Fig. \ref{cp}), 
falling slightly short (80$\%$) of $R$ln2 = 5.76 J/molK. 
No additional anomalies are observed up to 300 K (inset to Fig. \ref{cp}).

The triply degenerate $t_{2g}$ orbitals of OsO$_6$ octahedra in the undistorted Ba$_2$NaOsO$_6$ crystal structure 
constitute an effective unquenched angular momentum $\bf{L}$ = 1. 
The matrix elements of the orbital angular momentum operator $\bf{L}$ within the $t_{2g}$ manifold are the same as 
those of $-\bf{L}$ within states of $P$ symmetry \cite{Goodenough}. 
Application of spin orbit coupling therefore results in a quartet ground state ($J$ = 3/2)
and a doublet excited state ($J$ = 1/2) \cite{Goodenough, Stevens, Ishihara}.  
Since the integrated entropy through the magnetic phase transition is $\sim$ $R\ln2$ we can surmise 
that the quartet must be split into two Kramers doublets at a temperature $T_t$ $>$ $T_C$.

Xiang and Whangbo have recently suggested that both the insulating behavior and the ferromagnetism in Ba$_2$NaOsO$_6$ can be attributed to the 
combined effect of electron correlation and spin-orbit coupling \cite{Xiang}. However, the authors were unable to account for the magnitude of 
the saturated moment, the observed anisotropy, or the negative Weiss temperature. 
Furthermore, their assumption that the exchange energy is substantially larger than the spin-orbit coupling 
is not justified in light of the small value of $T_C$. These factors all 
indicate that a different mechanism is responsible for the ferromagnetism in this material. 

The quartet groundstate in Ba$_2$NaOsO$_6$ anticipated from consideration of the spin-orbit coupling is unstable to a Jahn-Teller distortion. 
Our observation of a doublet ground state certainly indicates 
that the anticipated degeneracy has been lifted, whilst the absence of any additional features in the heat capacity implies that $T_t$ $>$ 300 K. 
That this does not reveal itself in x-ray diffraction presumably reflects both the 
extremely subtle nature of the associated distortions (since the $t_{2g}$ orbitals are 
not elongated along the bond directions they couple more weakly to the oxygen ligands than do $e_g$ orbitals) 
and also the weak scattering power of the coordinating oxygen ions. 
The relative orientation of Os orbitals on adjacent sites will have a profound effect on the magnetically ordered state, 
via both the magnetic anisotropy and the superexchange \cite{Khomskii}, 
which was not considered in the previous calculations. 
The small negative Weiss temperature deduced from susceptibility 
measurements clearly indicates that Ba$_2$NaOsO$_6$ is not a simple ferromagnet 
but rather has a non-trivial magnetic structure, which can then be understood in terms of 
an orbitally ordered state with a non-zero ordering wave vector. 

If a scenario in which orbital order drives the ferromagnetism in Ba$_2$NaOsO$_6$ is indeed appropriate, 
then the fact that the isostructural, isoelectronic analog Ba$_2$LiOsO$_6$ is found to be antiferromagnetic \cite{zl02} 
indicates that the factors determining the orbital order in these two compounds are remarkably finely balanced. 
It is intriguing to think that a similarly complex interplay between spin and orbital degrees of freedom might be present 
in these 5$d^1$ Mott insulators as in their better-studied 3$d$ analogs, for which spin-orbit coupling has a much weaker effect. 
Resonant x-ray scattering experiments are in progress to directly address this possibility. 

\section*{Acknowledgments}
The authors thank R. M. White, Z. Islam, D. Mandrus and A. Sleight for useful conversations. 
This work is supported by the DOE, Office of
Basic Energy Sciences, under contract no. DE-AC02-76SF00515, 
and the NSF under grant no.s DMR-0134613 and DMR-0554796. 
The authors would also like to thank Prof. V. K. Pecharsky and Dr. Y. Mudryk for using their Rigaku PXRD. A part of this work was carried out at the Ames Laboratory, which is operated for the US Department of Energy by Iowa State University under Contract No. W-7405-ENG-82.


\begin{thebibliography}{99}
\bibitem{tolerance_factor}The tolerance factor calculated from ionic radii = 0.99, indicating that the structure is very stable.
\bibitem{walt1}The bare value for Os is -13.74 eV \protect\cite{walt}. 
Charge transfer between Os and O results in a shift in energy which can be calculated in a self-consistent manner. 
The shift in energy in E$_d$(Os) is given by $\Delta E$ = U$_d\Delta Z_d$, with U$_d$ = 3.4 eV \protect\cite{walt} 
and $\Delta Z_d$ = 0.76, from which E$_d$(Os) = -13.74 - (3.4 $\times$ 0.76) = -16.32 eV.
\bibitem{walt2}Since the Os $d$ orbitals are much smaller than the 4-atom O $p$-orbital cluster 
associated with one $t_{2g}$ molecular orbital, 
the dominant contributions to $U$ will come from the Os site. 
The antibonding orbital has only a weak polarization, 
characterized by $\alpha_P = 0.067$ \protect\cite{walt}, 
such that roughly half of the charge resides in the 
Os $d$-orbital and half in the 4-atom O $p$-orbital cluster.  
$U$ will therefore contain contributions from the self-energy of the two half electrons on the 
Os site, $\frac{1}{4} U_d$; 
the coulomb repulsion between the Os $\frac{1}{2} e^-$ and the $\frac{1}{2} e^-$ in the $p$-orbital cluster 
a distance $d$ away, $\frac{1}{4}e^2/d$; 
and the coulomb attraction between the Os $\frac{1}{2} e^-$ and the positive 
charge on the electron-deficient neighbor a distance $r$ away, $\frac{1}{2} e^2/r$, where $r\sim2\sqrt{2}d$.
The resulting total contribution $U$ = $\frac{1}{4} U_d$ + $\frac{1}{4}e^2/d$ + $\frac{1}{2} e^2/r = 3.32$ eV.
\end{thebibliography}
\end{document}